\documentclass[twocolumn,aps,prb,showpacs,preprintnumbers,amsmath,amssymb,psfrac]{revtex4}

\usepackage{epsfig}
\usepackage{graphicx}% Include figure files
\usepackage{dcolumn}% Align table columns on decimal point
\usepackage{bm}% bold math

\renewcommand{\dag}{^{\dagger}}
 
\newcommand{\la}{\langle}
\newcommand{\ra}{\rangle}
 
\newcommand{\al}{_{\alpha}}

\newcommand{\sx}{\sigma_x}

\newcommand{\sz}{\sigma_z}

\newcommand{\De}{\Delta}
\newcommand{\La}{\Lambda}

\newcommand{\dl}{\partial_\ell}

\def\gapp{\lower.35em\hbox{$\stackrel{\textstyle>}{\sim}$}}
\def\lapp{\lower.35em\hbox{$\stackrel{\textstyle<}{\sim}$}}

\begin{document}
%

%\draft
\title{
Universal asymptotic behavior in flow equations of dissipative systems
}
\author{Tobias Stauber}
\affiliation{Institut f\"ur Theoretische Physik, Ruprecht-Karls-Universit\"at Heidelberg, Philosophenweg 19, D-69120 Heidelberg, Germany
}
%\date{\today}

%%%%%%%%%%%%%%%%%%%%%%%%%%%%%%%%%%%%%%%%%%%%%%%%%%%%%%%%%%%%%%%%%%%%%%%%%%%%%
\begin{abstract}
Based on two dissipative models, universal asymptotic behavior of flow equations for Hamiltonians is found and discussed. Universal asymptotic behavior only depends on fundamental bath properties but not on initial system parameters, and the integro-differential equations possess an universal attractor. The asymptotic flow of the Hamiltonian can be characterized by a non-local differential equation which only depends on one parameter - independent of the dissipative system or truncation scheme. Since the fixed point Hamiltonian is trivial, the physical information is completely transferred to the transformation of the observables. This yields a more stable flow which is crucial for the numerical evaluation of correlation functions. Furthermore, the low energy behavior of correlation functions is determined analytically. The presented procedure can also be applied if relevant perturbations are present as is demonstrated by evaluating dynamical correlation functions for sub-Ohmic environments. It can further be generalized to other dissipative systems. 
\end{abstract}
%%%%%%%%%%%%%%%%%%%%%%%%%%%%%%%%%%%%%%%%%%%%%%%%%%%%%%%%%%%%%%%%%%%%%%%%%%%%%
%
%
%
%%%%%%%%%%%%%%%%%%%%%%%%%%%%%%%%%%%%%%%%%%%%%%%%%%%%%%%%%%%%%%%%%%%%%%%%%%%%%
%%%%%%%%%%%%%%%%%%%%%%%%%%%%%%%%%%%%%%%%%%%%%%%%%%%%%%%%%%%%%%%%%%%%%%%%%%%%%
%%%%%%%%%%%%%%%%%%%%%%%%%%%%%%%%%%%%%%%%%%%%%%%%%%%%%%%%%%%%%%%%%%%%%%%%%%%%%
%%%%%%%%%%%%%%%%%%%%%%%%%%%%%%%%%%%%%%%%%%%%%%%%%%%%%%%%%%%%%%%%%%%%%%%%%%%%%
%
\maketitle
\section{Introduction}
The flow equation method, introduced by G{\l}azek and Wilson \cite{Gla94} and independently by Wegner,\cite{Weg94} has become a powerful tool to analyze many-body systems in nuclear and solid-state physics.\cite{Weg01}
The method is based on the continuous mapping of a given Hamiltonian on to its unitarily equivalent representations. This induces a flow of the system parameters which is governed by the so-called flow equations. The continuous transformations can be characterized by the anti-Hermitian generator $\eta$ and the flow equations of the Hamiltonian $H$ are then given by the differential form $\dl H=[\eta,H]$, $\ell$ denoting the flow parameter. The objective is to reach a simple fixed point Hamiltonian for $\ell\to\infty$. 

The flow equations for many-body Hamiltonians are non-linear, integro-differential equations and in general analytically non-approachable. Yet, it is possible to analyze the asymptotic behavior as was done for the spin-boson model\cite{Keh96} as well as for an electron-phonon model.\cite{Len96} It was found that the asymptotic behavior of a given energy scale $\De$ generally leads to $\De\to\De^*+a\ell^{-1/2}$, with $\De^*\equiv\De(\ell=\infty)$ and $a$ denoting a dimensionless constant.\cite{FootEne} So far, $\eta$ has been chosen such that $\De^*$ was finite and the asymptotic fixed point Hamiltonian has thus been characterized by this renormalized energy scale. 

In this paper we want to apply the flow equation technique to dissipative systems. The general Hamiltonian then consists of a system Hamiltonian $H_S$ with energy scale $\Delta_0$, a bath Hamiltonian $H_B$, and the interaction term $H_{SB}$. Instead of integrating out the bath degrees of freedom and thus reducing the effective configuration space, the flow equation method performs a sequence of infinitesimal unitary transformations with the objective that the interaction between the system and the bath becomes zero, i.e. $H_{SB}\to0$. This is done systematically by first ``rotating away'' interaction terms which connect states with high energy differences. If $\De^*$ is finite, the interaction between the system and the environment is thus transformed away in the following way: first decoupling the high-energy modes of the bath, then proceeding with decoupling the low-energy modes and in the end decoupling nearly degenerate states around the renormalized energy scale $\De_r=\De^*\neq0$. Whereas in the first two regimes the coupling constants decrease exponentially as functions of $\ell$, the coupling constant belonging to the renormalized energy scale $\De^*$ follows an algebraic decay according to $\ell^{-1/2}$, see Ref. \onlinecite{Keh96}.  

In order to evaluate spectral properties within the flow equation approach, observables need to be subjected to the same sequence of unitary transformations. This leads to flow equations for observables and the observable flow follows the systematic decoupling procedure of the Hamiltonian, i.e. the correlation function is first built up for high and low energies and the intermediate energy regime around $\De^*$ is determined at last. 

Flow equations are more stable for $\ell$-values far away from degeneracies. This means that the regime which comprises most spectral weight is determined by flow equations which are the most unstable. In case of the spin-boson model, Kehrein and Mielke circumvent this problem by mapping the asymptotic flow of the spin-boson model onto the asymptotic flow of the exactly solvable dissipative harmonic oscillator.\cite{Keh97} They thus integrated the flow equations numerically up to some finite $\ell^*$ and then employed a conservation law which was derived for the exactly solvable model. Nevertheless, this mapping was restricted to a certain parameter regime\cite{FootKM} and might not be applied to other, more general dissipative models.    

In this paper, we will set up flow equations with a different asymptotic behavior. The crucial point is that the fixed point Hamiltonian $H^*\equiv H(\ell=\infty)$ does not contain any characteristic renormalized energy scale, i.e. $\De^*=0$ and $H^*$ is thus just given by the non-interacting bath. As a consequence, the system is monotonously decoupled from the bath, starting from high-energy modes and finishing with the low-energy modes. But it also implies that the fixed point Hamiltonian does not contain any other information but the ground-state energy. If the physical renormalized energy scale is denoted by $\De_r$, the flow equations do not yield $\De_r=\De^*$ as before. All physical information is transferred on to the observable flow. 

The asymptotic behavior of the Hamiltonian is passed on to the observable flow as before. Since the asymptotic flow equations deal with the decoupling of the low-energy modes, the asymptotic behavior of the observable will determine the low-energy regime of correlation functions. The low-energy regime of dissipative systems is generally determined by universal power-laws. We thus find the connection between universal flow equations and universal spectral properties. We will further set up a conservation law, based on asymptotic, scale-invariant spectral functions, which will allow us to halt the numerical integration after some finite value $\ell^*$. Like this, we also obtain analytical results for the low energy behavior of correlation functions. The procedure should not be restricted to specific models as before.     

There are other advantages of flow equations which lead to a trivial fixed point Hamiltonian only consisting of the free bath. First, one can easily include a logarithmic grid around $\omega=0$ to increase the numerical precision while decoupling almost degenerate states. Second, the spectrum of a dissipative system at low coupling\cite{FootBS} is generally that of the free bath, consisting of a ground-state and a gap-less continuum.\cite{Bac98} Unitary transformations do not alter the spectrum, but if the fixed point Hamiltonian consisted of the system with the renormalized energy scale $\De^*\neq0$ plus the free bath, the continuous spectrum of the bath would be superposed by the discrete spectrum of the system. Universal asymptotic behavior thus also guarantees the correct spectrum.
   
The paper is organized as follows. In Sec. II, we will treat the exactly solvable dissipative harmonic oscillator. We will first summarize the analytic solution of Kehrein and Mielke in a comprised form and then introduce and discuss universal asymptotic behavior. We explicitly show the connection between the conservation law obtained from the exact solution and an asymptotic, scale-invariant spectral function obtained from the universal behavior. In Sec. III, we will consider the spin-boson model for two different truncation schemes. In Sec. IIIA, we investigate flow equations stemming from a form-invariant truncation, i.e. no additional coupling terms are generated. This will allow a direct comparison with the results of the dissipative harmonic oscillator, but will yield rather poor results for dynamical correlation functions. In Sec. IIIB, we will therefore extend the truncation scheme and allow an additional coupling term to be generated. This approach will yield very good results for correlation functions within a large range of the parameter space including sub-Ohmic baths. The reader who is only interested in this non-trivial application, can directly go to Sec. IIIB, which is self-contained.

%
%%%%%%%%%%%%%%%%%%%%%%%%%%%%%%%%%%%%%%%%%%%%%%%%%%%%%%%%%%%%%%%%%%%%%%%%%%%%%
%%%%%%%%%%%%%%%%%%%%%%%%%%%%%%%%%%%%%%%%%%%%%%%%%%%%%%%%%%%%%%%%%%%%%%%%%%%%%
%%%%%%%%%%%%%%%%%%%%%%%%%%%%%%%%%%%%%%%%%%%%%%%%%%%%%%%%%%%%%%%%%%%%%%%%%%%%%
%%%%%%%%%%%%%%%%%%%%%%%%%%%%%%%%%%%%%%%%%%%%%%%%%%%%%%%%%%%%%%%%%%%%%%%%%%%%%
%
\section{Dissipative Harmonic Oscillator}
\label{SectionQuantumDissipativeHarmonicOscillator}

We will first discuss the harmonic oscillator coupled to an environment.
Following the seminal work by Caldeira and Leggett,\cite{Cal83} we will model the bath as a set of non-interacting harmonic oscillators with a dense spectrum. We will also introduce the interaction induced renormalization of the potential so that the Hamiltonian is bounded from below. 
The dissipative harmonic oscillator shall thus be described by the following Hamiltonian:
\begin{align}
\label{HOscillator}
    H&=\frac{p^2}{2}+\kappa v q^2+
     \sum_{\alpha}\Big(\frac{p_{\alpha}^2}{2}
        +\frac{1}{2}
        \omega_{\alpha}^2\big(x_{\alpha}-\frac{\lambda_{\alpha}}{\omega_{\alpha}^2}q\big)^2\Big)
\end{align}
The parameter $\kappa$ can be decomposed as $\kappa=V_0/(mq_0^2)$ where $V_0$ shall denote the potential energy scale, $m$ the mass of the particle, and $q_0$ a length scale such that $v$ is dimensionless.\cite{FootScales} We identify the frequency of the harmonic oscillator as $\sqrt{2\kappa v/m}$. The operators obey the canonical commutation relations which
read ($\hbar=1$)
\begin{align}
 \left[q,p\right]=i\quad,\quad\left[x_{\alpha},p_{\alpha^{\prime}}\right]=i
 \delta_{\alpha,\alpha'}\quad.
\end{align}

The system is exactly solvable and has been investigated by means of several techniques.\cite{Wei99} 
The exact solution of the model via the flow equation approach was first obtained by Kehrein and Mielke.\cite{Keh97} In Sec. \ref{AnaHarm}, we will recall the solution in a comprised form in order to introduce notations. In Sec. IIB, we will then discuss flow equations which exhibit universal asymptotic behavior. In Sec. \ref{NumHarm}, numerical results are presented.

\subsection{Analytical Results}
\label{AnaHarm}
\subsubsection{Hamiltonian flow}
In order to solve the dissipative harmonic oscillator via flow equations,\cite{Weg94}
the generator $\eta$ of the infinitesimal unitary transformation is chosen
to be
\begin{align}
\label{GeneralGenerator_HARM}
        \eta&=i\Big(q\sum_{\alpha}\eta_{\alpha}^qp_{\alpha}
        +p\sum_{\alpha}\eta_{\alpha}^px_{\alpha}+
        \sum_{\alpha,\alpha'}\eta_{\alpha,\alpha'}x_{\alpha}p_{\alpha^{\prime}}
        \Big)\\
	&\equiv\eta^q+\eta^p+\eta_B\quad.\notag
\end{align}
The first two terms follow from the canonical choice $\eta_c=[H,V]$ with the off-diagonal part $V=-q\sum\al\lambda\al x\al$, but with generalized parameters $\eta\al^q$ and $\eta\al^p$. They will be determined later. The last term $\eta^B$ is needed to cancel new interaction terms between different bath modes which are generated by the flow equations. The solvability of the model is due to the fact that this cancellation is possible and that therefore the flow equations close exactly.

The constants of the generator are now chosen such that the Hamiltonian remains form-invariant during the flow, i. e. no new interaction terms are generated. This procedure is not unique; a possible parametrization is given by   
\begin{align}
&\eta_{\alpha}^p=\lambda_{\alpha}
f(\omega_{\alpha},\ell)\quad,\quad\eta_{\alpha}^q=-\lambda_{\alpha}
f(\omega_{\alpha},\ell)\quad,\notag\\
&\eta_{\alpha,\alpha'}=-\frac{\lambda_{\alpha}\lambda_{\alpha'}}{
\omega_{\alpha}^2-
\omega_{\alpha'}^2}\big(f(\omega_{\alpha},\ell)+f(\omega_{\alpha'},\ell)\big)\quad,\notag
\end{align}
where $f(\omega\al,\ell)$ denotes an arbitrary function.   

The flow of the system is characterized by the renormalized potential 
\begin{align}
\label{RenormPot}
\tilde{v}(\ell)\equiv v(\ell)+\kappa^{-1}\sum\al\frac{\lambda\al^2(\ell)}{2\omega\al^2}\quad.
\end{align}
Defining the effective frequency $\De^2(\ell)\equiv2\kappa\tilde v(\ell)$ and the spectral coupling function
\begin{align}
J(\omega,\ell)=\frac{1}{\De(\ell)}\sum_{\alpha}\frac{\lambda_{\alpha}^2(\ell)}
{\omega_{\alpha}}\delta(\omega-\omega_{\alpha})\quad,
\end{align}
the flow equations $\dl H=[\eta,H]$ are given by the following coupled integro-differential equations:\cite{FootZT}
\begin{align}
\label{DiffvHarm}
&\partial_{\ell}\De(\ell)=\int_0^\infty d\omega
J(\omega,\ell)\omega f(\omega,\ell)\quad,\\
&\partial_{\ell}J(\omega,\ell)=
2J(\omega,\ell)\big(\omega^2-\De^2(\ell)\big)f(\omega,\ell)-J(\omega,\ell)\frac{\partial_{\ell}\De(\ell)}{\De(\ell)}\notag\\\label{DiffJHarm}
&-2\De(\ell)J(\omega,\ell)\int_0^\infty d\omega'
\frac{J(\omega',\ell)\omega'}{\omega^2-{\omega'}^2}F(\omega,\omega',\ell)\quad,
\end{align}
where we defined $F(\omega,\omega',\ell)\equiv f(\omega,\ell)+f(\omega',\ell)$. Notice that the renormalization of the bath modes $\omega\al$ was neglected since it vanishes in the thermodynamic limit which is a typical feature of dissipative systems.\cite{Keh97} Further, there is no mass renormalization in this approach, and we disregarded the renormalization of the ground-state energy. 

To solve these equations, Kehrein and Mielke introduced the function
\begin{align}
R(z,\ell)=\sum_{\alpha}\frac{\lambda_{\alpha}^2(\ell)}
{z-\omega_{\alpha}^2}\quad,
\end{align}
for which the following differential equation holds:
\begin{align}
\label{DiffR}
&\partial_{\ell}R(z,\ell)=-\partial_{\ell}\De^2(\ell)\\\notag
&+
2\big(z-\De^2(\ell)-R(z,\ell)\big)\sum_{\alpha}\frac{\lambda_{\alpha}^2(\ell)}
{z-\omega_{\alpha}^2}f(\omega_{\alpha},\ell)
\end{align}
The algebraic equation $z-\De^2(\ell)-R(z,\ell)=0$ solves the above differential equation for any $z$, and apparently also for any $f(\omega\al,\ell)$. With $z^*=\De^2(\ell=\infty)$ one imposes the boundary condition $R(z^*,\ell)\rightarrow0$ for $\ell\rightarrow\infty$ which guarantees that system and bath are decoupled for $\ell\rightarrow\infty$. This yields a
self-consistent equation for $\De^*\equiv\De(\ell=\infty)$ which can be solved for $\ell=0$. Generally it reads
\begin{align}
\label{FreqSelfCon}
{\De^*}^2=\De^2(\ell)+\De(\ell)\int_0^\infty d\omega
\frac{J(\omega,\ell)\omega}{{\De^*}^2-\omega^2}\quad.
\end{align} 

We already want to stress that for the above argumentation $\De^*$ has to be finite. Otherwise, Eq. (\ref{FreqSelfCon}) at $\ell=0$ and Eq. (\ref{RenormPot}) cannot hold simultaneously unless $v(\ell=0)=0$. The case $v(\ell=0)=0$ describes the system of a dissipative free particle. 

In the next subsection, we will choose $f(\omega\al,\ell)$ such that $\De^*=0$, and the system will still be decoupled from the bath for $\ell\to\infty$. The resulting flow equations thus belong to a different universality class since Eq. (\ref{FreqSelfCon}) does not hold anymore. This is the key observation of this subsection.

\subsubsection{Observable flow}

To determine correlation functions we have to apply the same sequence of
infinitesimal transformations to the observables that led to the
diagonalization of the Hamiltonian. Only then the diagonal structure of the Hamiltonian can be used to yield a simple time evolution of the operators in the Heisenberg picture. For the flow of the position operator we make the ansatz
\begin{align}
q(\ell)=h(\ell)q+\sum_{\alpha}\chi_{\alpha}(\ell)x_{\alpha}\quad.
\end{align}
The initial conditions are given by $h(\ell=0)=1$ and $\chi\al(\ell=0)=0$. During the flow, the weight of the system operator will be transferred to the bath operators. In the language of the flow equation approach we speak of a dissipative system when the total weight of the system is being transferred to the bath during the flow, i.e. $h(\ell=\infty)=0$.
The flow equations for the observable $\partial_{\ell}q=[\eta,q]$
close and with the spectral function
\begin{align}
	S(\omega,\ell)\equiv-\sum\al\frac{\lambda\al(\ell)\chi\al(\ell)}{\omega\al}
	\delta(\omega-\omega\al)
\end{align}
we obtain the following integro-differential equations:
\begin{align}
\label{DiffhHarm}
\dl h(\ell)=\int_0^\infty d\omega \omega S(\omega,\ell)f(\omega,\ell)\quad,
\end{align}
\begin{widetext}
\begin{align}
\notag
\dl S(\omega,\ell)&=-\De(\ell)h(\ell)J(\omega,\ell)f(\omega,\ell)
        -\De(\ell)J(\omega,\ell)\int_0^\infty d\omega' \frac{S(\omega',\ell)\omega'}
        {\omega^2-{\omega'}^2}F(\omega,\omega',\ell)\\
	&+(\omega^2-\De^2(\ell))S(\omega,\ell)f(\omega,\ell)
        -\De(\ell)
        S(\omega,\ell)\int_0^\infty d\omega' \frac{J(\omega',\ell)\omega'}
        {\omega^2-{\omega'}^2}F(\omega,\omega',\ell)\quad.\label{DiffchiHarm}
\end{align}
\end{widetext}

The flow equations for $p(\ell)=h(\ell)p+\sum_{\alpha}\chi_{\alpha}(\ell)p_{\alpha}$ are equivalent to Eqs. (\ref{DiffhHarm}) and (\ref{DiffchiHarm}) which merely demonstrates the fact that Hermite-city is conserved during the flow. From the above flow equations we obtain the following sum rule, which signifies that the commutation relation $[q(\ell),p(\ell)]=i$ holds for all $\ell$:
\begin{align}
h^2(\ell)+\sum_{\alpha}\chi_{\alpha}^2(\ell)=1
\end{align}
To solve the flow equations one introduces the functions
\begin{align}
\begin{split}
S_1(z,\ell)&=\sum_{\alpha}\frac{\lambda_{\alpha}\chi_{\alpha}}
{z-\omega_{\alpha}^2}\quad,\quad\\
S_2(z,\ell)&=\sum_{\alpha}\frac{\chi_{\alpha}^2}
{z-\omega_{\alpha}^2}\quad.
\end{split}
\end{align}
Kehrein and Mielke showed that the following quantity is conserved:
\begin{align}
\label{ConservationHarm}
S_2(z,\ell)+\frac{\big(h(\ell)-S_1(z,\ell)\big)^2}{z-\De(\ell)-R(z,\ell)}=\text{const.}
\end{align}
One can show that $h(\ell=\infty)=0$ for the initial conditions of interest such that the total weight of the system is indeed transferred to the bath.
Correlation functions are obtained through the following identity: 
\begin{align}
\notag
K(\omega,\ell)&\equiv\sum_{\alpha}\chi_{\alpha}^2(\ell)\delta(\omega^2-\omega_{\alpha}^2)\\\label{K_Harm}
&=\frac{1}{\pi}\text{Im}S_2(\omega^2-i0,\ell)
\end{align}
For example, the spectral function $\omega K(\omega)\equiv \omega K(\omega,\ell=\infty)$ is proportional to the Fourier transform of $\langle q(t)q\rangle$.

Choosing a Lorentzian spectral function
\begin{align}
\label{Lorentzian_Harm}
J(\omega)\equiv J(\omega,\ell=0)=\frac{4\gamma^2\omega\alpha}{\omega^2+\gamma^2}\quad,
\end{align}
the explicit solution is given by $(\omega_0\equiv\De(\ell=0))$\cite{Keh97}
\begin{widetext}
\begin{align}
\label{K_of_Lorentzian_Harm}
K(\omega)=\frac{2\alpha\gamma^2\omega_0\omega(\gamma^2+\omega^2)}
{(\omega_0^2(\gamma^2+\omega^2)-2\pi\alpha\gamma^3\omega_0-\omega^2(\gamma^2+\omega^2))^2+4\pi^2\alpha^2\omega_0^2\gamma^4\omega^2}\quad. 
\end{align}
\end{widetext}

\subsection{Universal Asymptotic Behavior}
\label{UniversalAsymptotics_HARM}

In the previous subsection, the generator of the flow equations, $\eta$, was chosen such that the initial Hamiltonian of Eq. (\ref{HOscillator}) remained form-invariant. This condition defines the generator only up to an arbitrary function $f(\omega,\ell)$. It is often useful to explicitly specify $f(\omega,\ell)$ even though the final result must be independent of the particular choice. In Ref. \onlinecite{Keh97}, Kehrein and Mielke choose $f(\omega,\ell)=-(\omega-\De(\ell))/(\omega+\De(\ell))$. This leads to $J(\omega,\ell)\propto\exp[-2(\omega-\De(\ell))^2\ell]$ for $\ell\to\infty$ if one neglects the non-linear term in Eq. (\ref{DiffJHarm}). The spectral function is therefore centered around the renormalized frequency $\De^*\neq0$. At $\omega=\De^*$ the spectral function vanishes algebraically as $\ell^{-1/2}$. The asymptotic spectral function thus depends on the initial frequency through $\De^*$. We want to label this asymptotic behavior non-universal.

A different choice is $f(\omega,\ell)=-1$ which would have the consequence that the renormalized potential has to tend to zero so that the system is decoupled for $\ell\to\infty$, i.e. $\De^*=0$ in order that $J(\omega,\ell=\infty)=0$. It therefore belongs to a different universality class since Eq. (\ref{FreqSelfCon}) does not hold anymore. The differential equation (\ref{DiffR}) turns into a Ricatti equation which can formally be integrated. 

That the choice $f(\omega,\ell)=-1$ really decouples the system from the bath is not clear from the beginning. One way to convince oneself is to study the asymptotic behavior of the flow equations (\ref{DiffvHarm}) and (\ref{DiffJHarm}). For this, we introduce an additional energy scale related to the coupling function, $\La(\ell)\equiv\int_0^\infty d\omega J(\omega,\ell)/\omega$.

We now make the ansatz 
\begin{align}
\label{PartHarmAsymp}
\De(\ell)\rightarrow a\ell^{-1/2}\quad,\quad
\La(\ell)\rightarrow b\ell^{-1/2}
\end{align}
with constants $a$ and $b$ as $\ell\to\infty$.
We further assume that\cite{FootDrop} 
\begin{align}
\label{SpectralHarmAsymp}
J(\omega,\ell)\rightarrow \omega^{s'}
\hat{J}(\ell)\bar{J}(\omega\sqrt\ell)\quad.
\end{align}

The various contributions of $J(\omega,\ell)$ can be interpreted as follows: The energy dependence $\omega^{s'}$ will yield the (correct) algebraic low-energy behavior of the spectral function, $\bar{J}(\omega\sqrt\ell)$ represents the high-energy cutoff function with the $\ell$-dependent cutoff frequency $\omega_c^\ell=\ell^{-1/2}$, and the function $\hat{J}(\ell)$ is needed to assure units of energy. 

The differential equations for $\La(\ell)$ and $\hat{J}(\ell)$ can be obtained from Eq. (\ref{DiffJHarm}) and the special feature of the exactly solvable dissipative harmonic oscillator is that the differential equation for $\La(\ell)$ closes, i.e. only $\La(\ell)$, $\De(\ell)$, and $\dl \De(\ell)$ are involved. One obtains $\hat{J}(\ell)\rightarrow \ell^{(s'-1)/2}$ and
\begin{align}
a^2=1/2+s'/4\quad,\quad b=a/2\quad.
\end{align}

We now define the dimensionless, scale-invariant variable $y\equiv\omega\sqrt{\ell}$ and set
$\bar{J}(y)\rightarrow J_0$ for $y\rightarrow0$. We then obtain the following non-local differential equation for 
$J(y)\equiv y^{s'-1}\bar{J}(y)$:
\begin{align}
\label{nonlocalJHarm}
\partial_y J(y)&=-4yJ(y)\Big(1-2a\int_0^{\infty}dy'\frac{J(y')}{y^2-{y'}^2}\Big)\\
\notag
&+(s'-1)\frac{J(y)}{y}
\end{align}
The boundary condition is given by $a=2\int_0^\infty dyJ(y)$.

The above ansatz for the asymptotic behavior (Eq. (\ref{PartHarmAsymp}) and (\ref{SpectralHarmAsymp})) guarantees that the system will be decoupled from the bath since the support of the spectral function vanishes as $\ell^{-1/2}$ and $J(y)\rightarrow y^{4+(s'-1)}e^{-2y^2}$ for $y\to\infty$. There is thus a universal fixed point for all initial frequencies. They are all mapped onto the free particle plus bath. 

To determine the asymptotic behavior of $h(\ell)$ and $S(\omega,\ell)$, we need the quantity $\Sigma(\ell)\equiv\int_0^\infty d\omega S(\omega,\ell)/\omega$. We make a similar ansatz as in the case of the spectral function, namely
\begin{align}
h(\ell)\rightarrow c\ell^{-1/2-\xi}\quad,\quad
\Sigma(\ell)\rightarrow d\ell^{-1/2-\xi}
\end{align}
with constants $c$ and $d$ as $\ell\to\infty$ and 
\begin{align}
S(\omega,\ell)\rightarrow \omega^{s'}
\hat{S}(\ell)\bar{S}(y)\quad.
\end{align}
Notice that we introduced a further parameter $\xi$.

The differential equation for $\Sigma(\ell)$ and $\hat{S}(\ell)$ are obtained from (\ref{DiffchiHarm}). One obtains $\hat{S}(\ell)\rightarrow \ell^{-\xi+(s'-1)/2}$ and 
\begin{align}
\xi=s'/4\quad,\quad c=2d\quad. 
\end{align}
Setting $\bar{S}(y)\rightarrow S_0$ for $y\rightarrow0$, we obtain the following non-local differential equation for $S(y)\equiv y^{s'-1}\bar{S}(y)$:
\begin{align}
\label{nonlocalSHarm}
\partial_y S(y)&=-2yS(y)\Big(1-2a\int_0^{\infty}dy'\frac{J(y')}{y^2-{y'}^2}\Big)\\
	&+4ayJ(y)\int_0^{\infty}dy'\frac{S(y')}{y^2-{y'}^2}\notag
	+(s'-1)\frac{S(y)}{y}
\end{align}
Eq. (\ref{nonlocalSHarm}) is linear in $S(y)$. We can therefore not determine the constant $S_0$ from the asymptotic behavior. But we can extract the low-frequency behavior of the correlation function $K(\omega)=\lim_{\ell\to\infty}S^2(\omega,\ell)/(2\De(\ell)J(\omega,\ell))$: 
\begin{align}
K(\omega\to0)=\omega^{s'}\frac{S_0^2}{2aJ_0}
\end{align}
The numerical results yield $s'=s$, where $s$ denotes the algebraic behavior of the initial spectral function at low energies, i.e. we recover the well-known result. 

Finally, we will give an alternative derivation of the conservation law of Eq. (\ref{ConservationHarm}) which is valid in the asymptotic regime. For this, we define the asymptotic function $K_a(\omega,\ell)$, following from the conservation law:
\begin{align}
K_a(\omega,\ell)\equiv \frac{1}{\pi}\text{Im}\frac{(h(\ell)-S_1(\omega^2-i0,\ell))^2}{\omega^2-\De^2(\ell)-R(\omega^2-i0,\ell)}
\end{align}
Taking $\ell^*$ to be in the asymptotic regime, we obtain $K_a(\omega,\ell)\equiv K(\omega)-K(\omega,\ell^*)=\omega^{s'}K_a(y^*)$ with $y^*=\omega\sqrt{\ell^*}$ and 
\begin{align}
\label{K_a_Asmptotic}
K_a(y^*)=4\int_{y^*}^{\infty}
dyy^{2-s'}S(y)\int_0^\infty dy'\frac{S(y')}{y^2-{y'}^2}\quad. 
\end{align}

One can thus derive the conservation law from the universal asymptotic behavior, given that $\ell^*$ is in the asymptotic regime. This is an important observation since conservation laws as Eq. (\ref{ConservationHarm}) can only be derived for solvable models. But by analyzing the asymptotic behavior, we are able to integrate the flow equations up to a finite but asymptotic $\ell^*$ and than complete the spectral function by analyzing the non-local differential equations for the scale-invariant spectral functions. This procedure also works for non-trivial models. 
 
\subsection{Numerical Results}
\label{NumHarm}
\begin{figure}[t]
  \begin{center}
    \epsfig{file=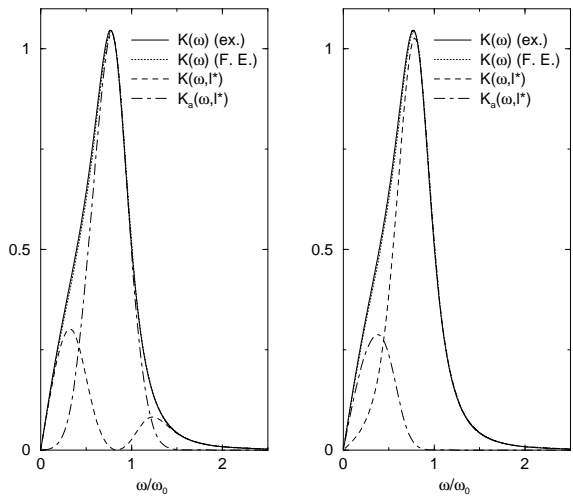,height=7.5cm,angle=-90}
    \caption{The spectral function $K(\omega)=K(\omega,\ell^*)+K_a(\omega,\ell^*)$ calculated via flow equations with $f(\omega,\ell)=-(\omega-\De(\ell))/(\omega+\De(\ell))$ (lhs) and with $f(\omega,\ell)=-1$ (rhs), taken at $\omega_0^2\ell^*=10$ for $J(\omega)=4\gamma^2\omega\alpha/(\omega^2+\gamma^2)$ with $\omega_0\equiv\De(\ell=0)$, $\alpha=0.1$ and $\gamma/\omega_0=1$ (dotted line). The solid line resembles the analytic solution of Eq. (\ref{K_of_Lorentzian_Harm}) which is almost entirely superposing the dotted line.}
    \label{Spectral_Harm}
\end{center}
\end{figure}
\begin{figure}[t]
  \begin{center}
    \epsfig{file=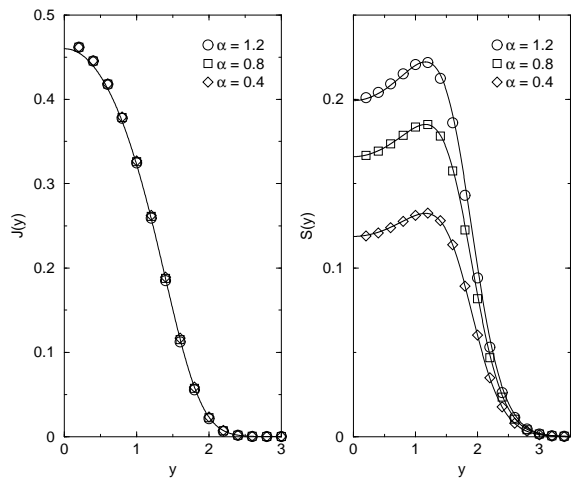,height=7.5cm,angle=-90}
    \caption{The asymptotic functions $J(y)$ and $S(y)$
at $v_0\ell^*=100$ for Ohmic coupling $J(\omega)=2\alpha\omega\Theta(\omega_c-\omega)$ with $\omega_c^2/v_0=100$ and $v_0\equiv \kappa v(\ell=0)$ for different coupling strengths $\alpha$. The solid lines resemble the ``analytic'' solution.}
    \label{Asymp_Harm}
\end{center}
\end{figure}
\begin{figure}[t]
  \begin{center}
    \epsfig{file=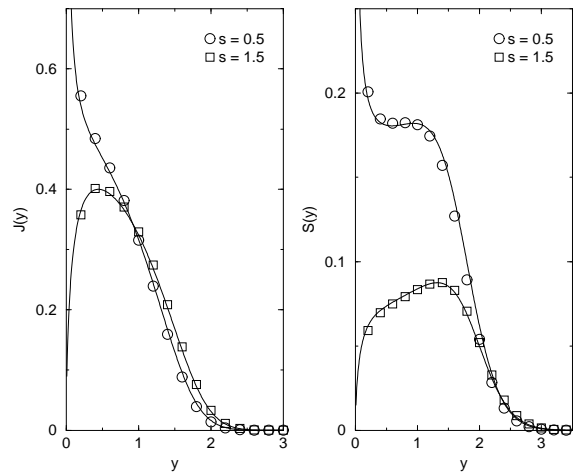,height=7.5cm,angle=-90}
    \caption{The asymptotic functions $J(y)$ and $S(y)$ with $s'=1$
at $v_0\ell^*=100$ for $J(\omega)=2\alpha K^{1-s}\omega^s\Theta(\omega_c-\omega)$ with $\alpha=0.4$, $K^2/v_0=1$, $\omega_c^2/v_0=100$ and $v_0\equiv \kappa v(\ell=0)$ for different bath types $s$ ($s'=s$). The solid lines resemble the ``analytic'' solution.}
    \label{Asymp_s}
\end{center}
\end{figure}
\begin{figure}[t]
  \begin{center}
    \epsfig{file=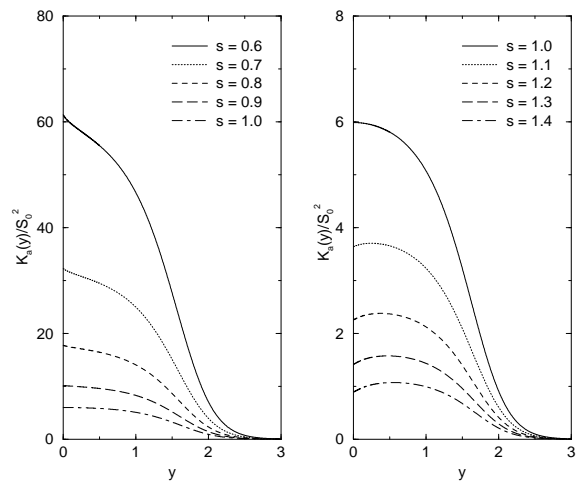,height=7.5cm,angle=-90}
    \caption{The universal asymptotic function $K_a(y)/S_0^2$ for sub-Ohmic (lhs), Ohmic, and super-Ohmic baths (rhs) ($s'=s$).}
    \label{Abschluss_s}
\end{center}
\end{figure}

We will now analyze the above flow equations numerically. Given the Lorentzian of Eq. (\ref{Lorentzian_Harm}) as initial coupling function, we will calculate $K(\omega)$ within the flow equation approach for two different choices of $f(\omega,\ell)$. To do so, we will employ the conservation law of Eq. (\ref{ConservationHarm}), i.e. $K(\omega)=K(\omega,\ell)+K_a(\omega,\ell)$. This allows us to halt the integration routine after a finite $\ell^*$. Nevertheless, the final result must be independent of $\ell^*$ which is the case.

In Fig. \ref{Spectral_Harm}, the results are shown for $\alpha=0.1$ and $\gamma/\omega_0=1$. The analytic solution superposes the solution obtained by flow equation approach and, obviously, $K(\omega)=K(\omega,\ell)+K_a(\omega,\ell)$ is independent of the two choices of $f(\omega,\ell)$. Nevertheless, the functions $K(\omega,\ell)$ and $K_a(\omega,\ell)$ are different. For $f(\omega,\ell)=-(\omega-\De(\ell))/(\omega+\De(\ell))$, $K_a(\omega,\ell)$ mostly contains the weight around intermediate time scales, whereas for $f(\omega,\ell)=-1$, $K_a(\omega,\ell)$ determines the low-frequency behavior.\\

We will now investigate the asymptotic behavior of the flow equations for $f(\omega,\ell)=-1$.
The differential equations (\ref{nonlocalJHarm}) and (\ref{nonlocalSHarm}) can be solved numerically via self-consistent iteration. Comparing these ``analytic'' results with the numerical results obtained via integrating the flow equations shall serve as a proof that we have found the correct  asymptotic behavior.

In the following, the initial spectral coupling function shall be given by $J(\omega)=2\alpha K^{1-s}\omega^s\Theta(\omega_c-\omega)$ where $\alpha$ denotes the coupling strength and $\omega_c$ the high-frequency cutoff. The bath type is labeled by $s$ and $K$ denotes an additional energy scale present for $s\neq1$. One distinguishes the following three characteristic coupling types: super-Ohmic coupling ($s>1$), Ohmic coupling $(s=1)$, and sub-Ohmic coupling ($s<1$). 
The numerical data shows that $s'=s$.

The results for Ohmic coupling $s=1$ are shown in Fig. \ref{Asymp_Harm}, where the asymptotic curves $J(y)$ and $S(y)$ obtained through direct integration and through self-consistent iteration are plotted for different coupling strengths $\alpha$. Indeed, $J(y)$ is universal for all coupling strengths, the functions $S(y)$ only differ by the constant $S_0$. This behavior also holds for $s\neq1$. 

In Fig. \ref{Asymp_s}, the scale-invariant functions $J(y)$ and $S(y)$ are shown at fixed coupling strength $\alpha=0.4$ for two different parameters $s$. In the super-Ohmic case $s=1.5$, $J(y)$ and $S(y)$ show a suppression whereas in the sub-Ohmic case $s=0.5$, $J(y)$ and $S(y)$ diverge as $y\to0$.

We close this section with a discussion on the asymptotic function $K_a(y)$ of of Eq. (\ref{K_a_Asmptotic}) which is needed for the alternative conservation law. $K_a(y)$ is proportional to $S_0^2$ and thus depends on the coupling strength $\alpha$. Scaling out this factor, we obtain universal functions only depending on the bath type $(s'=s)$. In Fig. \ref{Abschluss_s}, the results are shown for sub-Ohmic (lhs), Ohmic, and super-Ohmic baths (rhs). Notice that only for Ohmic dissipation $s=1$, $\partial_y K(y)\to0$ for $y\to0$. 

\section{Spin-Boson Model}

The most prominent dissipative quantum system is the spin-boson model. It has been extensively studied over the past thirty years by several varying techniques - ranging from the Feynman-Vernon influence-functional formulation,\cite{Fey65} Liouville operator and projection methods,\cite{Fic90} to exact mathematical results.\cite{Bac95} Its popularity stems from the fact that it is the most simple non-trivial dissipative quantum system, consisting only of a two-state system and a bosonic bath that are linearly coupled in order to preserve the reflection symmetry of the system. Nevertheless, applications for the model are found in all fields of physics, starting from quantum optics and solid state physics to nuclear physics and chemistry.\cite{Wei99,Leg87} In addition, from a mathematical and theoretical point of view, it attracts a lot of interest since it exhibits a Kosterlitz-Thouless transition for Ohmic coupling at $T=0$ and a crossover from coherent to incoherent tunneling at finite temperature for all coupling types. Moreover, it is related to other prominent models of theoretical physics - most strikingly to the anisotropic Kondo model. 

The Kondo model as well as the spin-boson model belong to the so-called strong-coupling problems which make the use of renormalization group techniques almost indispensable. The low-energy behavior of the Kondo model was thus first obtained by Wilson employing his numerical renormalization group.\cite{Wil75} These ideas were also applied to the spin-boson model.\cite{Cos96} Flow equations results for the spin-boson model and Kondo model can be found in Refs. \onlinecite{Keh97, Hof01, Sle03, Sta02}. 

The Hamiltonian of the symmetric spin-boson model without bias is given by 
\begin{align}
\label{Hamiltonian_SpinBoson_Initial}
  \begin{split}
    H=-\frac{\De_0}{2}\sigma_x + \sum_{\alpha}
        \omega_{\alpha}b_{\alpha}\dag b_{\alpha}
     +\sigma_z\sum_{\alpha}\frac{\lambda_{\alpha}^0}{2}
        (b_{\alpha}+b_{\alpha}\dag).
  \end{split}
\end{align}

The operators $b_{\alpha}^{(\dagger)}$ resemble the bath degrees of freedom
and $\sigma_i$ with $i=x,y,z$ denote the Pauli spin matrices. They
obey the canonical commutation relations and the spin-$1/2$ algebra, respectively. The coupling constants $\lambda\al$ only enter via the spectral coupling function 
\begin{align}
J(\omega)=\sum_{\alpha}(\lambda_{\alpha}^0)^2
\delta(\omega-\omega_{\alpha}),
\end{align}
which shall be parametrized as  
\begin{align}
J(\omega)=2\alpha K^{1-s}\omega^s\Theta(\omega_c-\omega).
\end{align}
The dimensionless parameter $\alpha$ denotes the coupling strength and $\omega_c$ the high-frequency cutoff. The bath type is labeled by $s$ and $K$ denotes an additional energy scale present for $s\neq1$.

The Hamiltonian (\ref{Hamiltonian_SpinBoson_Initial}) resembles an effective Hamiltonian where the high energy degrees of freedom of the bath were already integrated out - down to the bath cutoff $\omega_c$ by employing the Born-Oppenheimer approximation. The tunnel-matrix element $\De_0$ thus depends on this arbitrary energy scale and all physical results must be independent of the cutoff $\omega_c$. For Ohmic coupling one obtains $\Delta_0\propto(\omega_c)^\alpha$ and the only combination of $\Delta_0$ and $\omega_c$ that yields the units of an energy and is independent of $\omega_c$ is given by $\Delta_{r}\propto\Delta_0(\Delta_0/\omega_c)^{\alpha/(1-\alpha)}$. Physical observables should only depend on this effective tunnel-matrix element.

In the following, we will discuss the symmetrized equilibrium correlation function 
\begin{align}
C(t)=\frac{1}{2}\la\{\sz(t),\sz\}\ra =\int_0^\infty dt e^{i\omega t}C(\omega)
\end{align}
at low coupling. In this regime, $C(\omega)$ has been evaluated for super-Ohmic baths and for Ohmic baths by Kehrein and Mielke employing flow equations which exhibited non-universal asymptotic behavior.\cite{Keh97} Here, we will set up and investigate flow equations which exhibit universal asymptotic behavior.

In Sec. \ref{FormInv}, we first discuss flow equations which preserve the form of the Hamiltonian of Eq. (\ref{Hamiltonian_SpinBoson_Initial}). A direct comparison with the flow of the dissipative harmonic oscillator is then possible, but this truncation scheme will yield rather poor results for $C(\omega)$. In Sec. \ref{Extended}, we will therefore extend the truncation scheme and also allow additional coupling terms to be generated. This procedure will yield very good results for equilibrium correlation functions even for sub-Ohmic baths. 

This extended scheme includes operators which we want to label ``transitionally relevant''. Those operators neither appear in the initial nor in the fixed point Hamiltonian, but are obviously needed as transition link. This extents the classification of operators for the flow equation approach, initiated by Kehrein and Mielke in Ref. \onlinecite{Keh94}. We want to note that this ``transitional relevance'' can already be seen from a simpler model including only one bosonic mode.\cite{Sta03}

\subsection{Form-invariant flow}
\label{FormInv}

\subsubsection{Flow equations}
Applying flow equations to the spin-boson model we will first be guided by the premise that the flow shall not generate new interaction terms. This cannot be done exactly as in the case of the dissipative harmonic oscillator but one has to neglect normal ordered coupling terms of higher order. Then, one can follow the same procedure as outlined in the previous section. We therefore just cite the resulting equations. For details, we refer to Ref. \onlinecite{Keh97} and Ref. \onlinecite{StaD}.  

The flow equations $\dl H=[\eta,H]$ for the spin-boson model with form-invariant flow are given by the following coupled integro-differential equations ($T=0$):
\begin{align}
\dl \De(\ell) &=\De(\ell) \int_0^\infty d\omega
J(\omega,\ell)f(\omega,\ell)\\
\dl J(\omega,\ell)&=
2J(\omega,\ell)(\omega^2-\De^2(\ell))f(\omega,\ell)\\\notag
&-2\De(\ell) J(\omega,\ell)\int_0^\infty d\omega'
\frac{J(\omega',\ell)\omega'}{\omega^2-{\omega'}^2}F(\omega,\omega',\ell)
\end{align}
We introduced the $\ell$-dependent spectral coupling function $J(\omega,\ell)\equiv\sum\al\lambda\al^2(\ell)\delta(\omega-\omega\al)$ and defined $F(\omega,\omega',\ell)=f(\omega,\ell)+f(\omega',\ell)$. Further, we disregarded the renormalization of the ground-state energy.

As in the previous section, the condition of form-invariance does not specify the flow completely and the flow equations still depend on an arbitrary function $f(\omega,\ell)$. But if one considers the spin-boson model with finite bias or starts from a unitarily equivalent representation of the initial Hamiltonian where the reflection symmetry is broken, e.g. $b\al\to b\al -\frac{\lambda\al^0}{2\omega\al}$, there is no open choice for $f(\omega,\ell)$ anymore and $f(\omega,\ell)=-1$ emerges automatically.\cite{StaD} Breaking the reflection symmetry of the system thus also breaks the ``gauge-invariance'', expressed by the function $f(\omega,\ell)$.

The truncated flow of the observable shall be given by 
\begin{align}
\sigma_z(\ell)=h(\ell)\sigma_z+\sigma_x\sum\al\chi\al(\ell)(b\al+b\al\dag)\quad.
\end{align}
The flow equations for the observable $\partial_\ell\sigma_z=[\eta,\sigma_z]$ are approximated to yield
\begin{align}
\dl h(\ell)&=\De(\ell)\int_0^\infty d\omega S(\omega,\ell)f(\omega,\ell)\quad,
\end{align}
\begin{widetext}
\begin{align}
\label{diffS}
\dl S(\omega,\ell)&=-\De(\ell) h(\ell)J(\omega,\ell)f(\omega,\ell)
        -\De(\ell)
        J(\omega,\ell)\int_0^\infty d\omega' \frac{S(\omega',\ell)\omega'}
        {\omega^2-{\omega'}^2}F(\omega,\omega',\ell)\\\notag
	&+(\omega^2-\De^2(\ell))S(\omega,\ell)f(\omega,\ell)
        -\De(\ell)
        S(\omega,\ell)\int_0^\infty d\omega' \frac{J(\omega',\ell)\omega'}
        {\omega^2-{\omega'}^2}F(\omega,\omega',\ell)\quad,
\end{align}
\end{widetext}
where we defined $S(\omega,\ell)\equiv\sum\al\lambda\al(\ell)\chi\al(\ell)\delta(\omega-\omega\al)$.

In Ref. \onlinecite{Keh97}, the flow equations are numerically integrated,
choosing $f(\omega,\ell)=-(\omega-\De(\ell))/(\omega+\De(\ell))$. 
The asymptotic spectral function is then 
centered around the renormalized tunnel-matrix element $\Delta_r$ and one is able to map the asymptotic flow equations of the
spin-boson model on to the asymptotic flow equations of the dissipative
harmonic oscillator. One can then employ the exact solution of the
dissipative harmonic oscillator outlined in Sec. IIA. The result is shown in Fig. \ref{FinalFinalFig}.

\subsubsection{Asymptotic behavior}

In analogy to the dissipative harmonic oscillator we now investigate the flow equations with $f(\omega,\ell)=-1$ which will lead to universal asymptotic behavior. These flow equations show apparent deficiencies when calculating correlation functions. E.g., the sum rule $\langle\sz^2(\ell)\rangle=1$ is not satisfied for $f(\omega,\ell)=-1$ which can be seen from the differential equation of $\langle\sz^2(\ell)\rangle$,
\begin{align}
\dl \langle\sz^2(\ell)\rangle=-2\De(\ell)\int_0^\infty d\omega
d\omega'\frac{S(\omega,\ell)S(\omega',\ell)}{\omega+\omega'}\quad.
\end{align}
Since $h(\ell)\to0$ for $\ell\to\infty$ and $S(\omega,\ell)\geq0$, the spectral function $C(\omega)=\sum\al\chi\al^2(\ell=\infty)\delta(\omega-\omega\al)$ cannot satisfy the spectral sum rule $\int_0^\infty d\omega C(\omega)=1$. 

Analyzing the asymptotic behavior, we now show that the flow equations do not yield the right asymptotic behavior, either. We limit ourselves to Ohmic dissipation. 

Following the discussion of the previous section, we make the ansatz 
\begin{align}
\De(\ell)\rightarrow a\ell^{-1/2}\quad,\quad
\La(\ell)\rightarrow b\ell^{-1/2} 
\end{align}
with $\La(\ell)\equiv\int_0^\infty d\omega J(\omega,\ell)/\omega$ as $\ell\to\infty$.
We further assume that $J(\omega,\ell)\rightarrow \omega J(y)$, where $J(y)\to J_0$ for $y\to0$ and $y\equiv\omega\sqrt{\ell}$. We observe that $a=2b$ and
thus obtain the following non-local differential equation:
\begin{align}
\label{nonlocalJ}
\partial_y J(y)=-4yJ(y)\Big(1-2a\int_0^{\infty}dy'\frac{J(y')}{y^2-{y'}^2}\Big)
\end{align}
The boundary condition is given by $1=2\int_0^\infty dyyJ(y)$.

The above equation is identical to Eq. (\ref{nonlocalJHarm}) with the only difference that the parameter $a=2\int_0^\infty dyJ(y)$ takes on a different value and cannot be determined analytically, now. In the next subsection, we will see that the above equation also describes the asymptotic flow of the coupling function for an improved truncation scheme, but with $a=2$. 

To determine the asymptotic behavior of $S(\omega,\ell)$ we make the ansatz 
\begin{align}
h(\ell)\rightarrow c\ell^{-1/2-\xi}\quad,\quad
\Sigma(\ell)\rightarrow d\ell^{-1/2-\xi}
\end{align}
with $\Sigma(\ell)\equiv\int_0^\infty d\omega S(\omega,\ell)/\omega$ as $\ell\to\infty$. Further we assume that $S(\omega,\ell)\rightarrow \omega
\hat{S}(\ell)S(y)$, where $S(y)\rightarrow S_0$ for $y\rightarrow0$.
This leads to the following non-local differential equation for $S(y)$:
\begin{align}
\label{nonlocalS}
\partial_y S(y)&=-2yS(y)\Big(1-2a\int_0^{\infty}dy'\frac{J(y')}{y^2-{y'}^2}\Big)\\
	+4ay&J(y)\int_0^{\infty}dy'\frac{S(y')}{y^2-{y'}^2}\notag
	+2\xi\frac{S(y)-J(y)S_0/J_0}{y}
\end{align}
There is one difference to the corresponding differential equation of the dissipative harmonic oscillator since the last term in Eq. (\ref{nonlocalS}) is not present in Eq. (\ref{nonlocalSHarm}). 

To see the deficiency of the above asymptotic behavior, we now observe that
\begin{align}
C(\omega,\ell)=\frac{S^2(\omega,\ell)}{J(\omega,\ell)}\to\omega\ell^{-2\xi}\frac{S_0^2}{J_0}\quad. 
\end{align}
We would only obtain the correct asymptotic behavior $C(\omega)\propto\omega$ for $\omega\to0$ if $\xi=0$. As we will see from the numerical results, this is not the case. We thus obtain the wrong asymptotic behavior which is an evidence for the general shortcoming of the form-invariant flow.

We want to close with a remark on non-universal asymptotic behavior. For Ohmic coupling it yields $\De(\ell)\to \De^*+\frac{1}{2}\ell^{-1/2}$ for coupling constants $\alpha<1$, with $\De^*$ denoting the renormalized tunnel-matrix element $\De^*\propto\Delta_0(\Delta_0/\omega_c)^{\alpha/(1-\alpha)}$, see Ref. \onlinecite{Keh96}. For $\alpha>1$ the asymptotic behavior is governed by $\Delta\to\ell^{-1/2}/\ln\ell$. The localization phenomena can thus be detected by the different asymptotic behavior of $\De(\ell)$. This is in contrast to the universal asymptotic behavior where for all coupling constants $\De(\ell)\to a\ell^{-1/2}$ holds. This indicates that our approach will only be valid for small coupling strength, away from the broken symmetry phase.

\subsubsection{Numerical results}

\begin{figure}[t]
  \begin{center}
    \epsfig{file=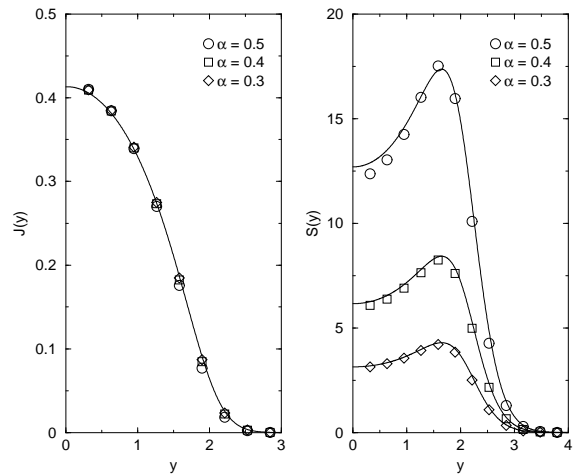,height=7.5cm,angle=-90}
    \caption{The asymptotic spectral functions $J(y)$ and $S(y)$ obtained from the form-invariant flow for Ohmic coupling at $\Delta_0^2\ell^*=10^3$ for different coupling constants $\alpha$. The solid line resembles the ``analytic'' solution.}
    \label{Spektralplus_SPINBOSON}
  \end{center}
\end{figure}
\begin{figure}[t]
  \begin{center}
    \epsfig{file=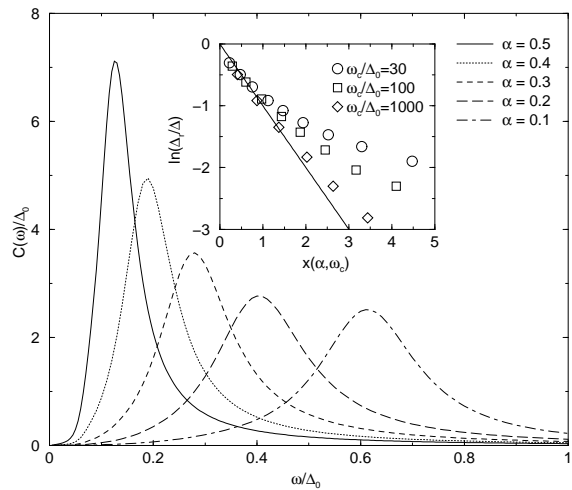,height=7.5cm,angle=-90}
    \caption{The spectral functions $C(\omega)$ obtained from the form-invariant flow for Ohmic coupling $J(\omega)=2\alpha\omega\Theta(\omega_c-\omega)$ with $\omega_c/\De_0=30$ at $\Delta_0^2\ell^*=10^3$ for different coupling constants $\alpha$. Inset: Defining the maximum of $C(\omega)$ as $\De_r$, $\ln(\De_r/\De_0)$ is plotted versus $x(\alpha,\omega_c)\equiv\alpha/(1-\alpha)\ln(\omega_c/\De_0)-\ln(\cos(\pi\alpha)\Gamma(1-2\alpha))/(2-2\alpha)$ for different cutoffs $\omega_c$. The solid line resembles the NIBA result for $\De_{\text{eff}}$.}
    \label{SpectralUnivSpBo}
  \end{center}
\end{figure}

The predictions of the above ansatz can be verified numerically by integrating the flow equations. In contrast to the exactly solvable dissipative harmonic oscillator, the constants $a$ and $\xi$ cannot be determined analytically. The self-consistent solution of the differential equations (\ref{nonlocalJ}) and (\ref{nonlocalS}) yields the following values: 
\begin{align}
a\approx1.21\quad,\quad\xi\approx 0.41
\end{align}

In Fig. \ref{Spektralplus_SPINBOSON}, the universal asymptotic functions $J(y)$ and $S(y)$ obtained from numerical integration of the flow equations are compared with the ``analytic'' solution following from the self-consistent evaluation of Eqs. (\ref{nonlocalJ}) and (\ref{nonlocalS}). The asymptotic results do only depend on the bath type (not shown here), but not on the coupling strength $\alpha$, the explicit form of the cutoff function, or on the value of $\omega_c$.

The symmetrized equilibrium correlation function $C(\omega)$ must of course depend on the initial conditions.
Since the asymptotic functions $S(y)$ only differ by at most a constant for arbitrary initial conditions, this information must be contained in the primary and intermediate $\ell$-range of the flow. In Fig. \ref{SpectralUnivSpBo}, the flow equation results of $C(\omega)$ for a Ohmic bath with sharp cutoff at $\omega_c/\De_0=30$ at different coupling strengths $\alpha$ are shown.

The spectral function $C(\omega)$ is obtained by integrating the flow equations up to $\De_0^2\ell^*=10^3$ where one is already in the asymptotic regime for the coupling strengths considered. For higher coupling strengths one would have to integrate up to larger $\ell^*$, since $\De_r$ is smaller. Notice that we do not employ a conservation law to build up the spectral functions, nor is our approach limited to small coupling strengths. But one crucial shortcoming of the flow equations is that $C(\omega)$ does not fulfill the sum rule $\int_0^\infty d\omega C(\omega)=1$. Independent from the cutoff parameter, $\omega_c$ and bath type the sum rule only yields approximately $80\%$ for small coupling constants ($\alpha\lapp0.1$) and $83-85\%$ for larger coupling constants.

Nevertheless, $C(\omega)$ contains physical information. This is demonstrated in the inset of Fig. \ref{SpectralUnivSpBo} where the maximum of $C(\omega)$ is plotted for different cutoff frequencies $\omega_c$ as a function of the coupling strength $\alpha$, according to the effective tunnel-matrix element 
\begin{align}
\Delta_{\text{eff}}\equiv\big(\cos(\pi\alpha)\Gamma(1-2\alpha)\big)^{-2(1-\alpha)}\Delta_0\Big(\frac{\Delta_0}{\omega_c}\Big)^{\frac{\alpha}{1-\alpha}},
\end{align} 
derived from the non-interacting blib approximation (NIBA).\cite{Leg87} The universal regime described by the NIBA is only reached for small coupling constants and for rather large $\omega_c$. Still, the flow equations capture the essential physics despite of the apparent shortcomings of the correlation function.

\vspace{1cm}
\subsection{Extended flow equations}
\label{Extended}
\subsubsection{Flow equations}

Since the flow equations of the form-invariant flow did not yield satisfactory results, we will now extend the truncation scheme, i.e. we will choose a different generator and we will allow a new coupling term to be generated. But universal asymptotic behavior shall be retained.

The anti-Hermitian generator $\eta$ shall first be chosen canonically, i.e. $\eta_c=[H_B,H]$, with the diagonal Hamiltonian $H_B=\sum\al\omega_{\alpha}b_{\alpha}\dag b\al$ and $H$ given in Eq. (\ref{Hamiltonian_SpinBoson_Initial}). This choice indicates that there will be universal asymptotic behavior since the diagonal Hamiltonian consists of the bath only and is thus $\ell$-independent.\cite{FootFS} 

The commutator $[\eta_c,H]$ will now generate a new coupling term which is linear in the bosonic operators. Neglecting the shift in the ground-state energy, we thus arrive at the following truncated Hamiltonian:
\begin{align}
H(\ell)&=-\frac{\De(\ell)}{2}\sigma_x+\sum\al\omega\al b\al\dag b\al\\\notag
&+i\sigma_y\sum\al\frac{\lambda\al^y(\ell)}{2}(b\al-b\al\dag)
+\sigma_z\sum\al\frac{\lambda\al^z(\ell)}{2}(b\al+b\al\dag)
\end{align}
with $\De(\ell=0)=\De_0$, $\lambda\al^z(\ell=0)=\lambda\al^0$, and $\lambda\al^y(\ell=0)=0$.

The canonical generator $\eta_c$ also gives rise to coupling terms which are bilinear in the bosonic operators. These terms will not be included in the Hamiltonian flow explicitly. Still, they are adequately taken into account of by introducing an additional term to the generator which approximately cancels these bilinear contributions. This is in analogy to the procedure of the previous section. For details, see Ref. \onlinecite{Sta02}. 

In terms of the spectral coupling functions $J^k(\omega,\ell)\equiv\sum\al(\lambda\al^k)^2(\ell)\delta(\omega-\omega\al)$, with $k=y,z$, we obtain the following flow equations:
\begin{align}
\dl \De &=-2\int_0^\infty  d\omega \omega J^m(\omega,\ell)\\
\label{FlowEquations_Extended_SpinBoson}
\dl J^k(\omega,\ell)&=-2\omega^2J^{\bar k}(\omega,\ell)+2\De\omega J^m(\omega,\ell)\\\notag
&+2J^k(\omega,\ell)\int_0^\infty  d\omega'J^m(\omega',\ell)
\frac{\omega^2+{\omega'}^2}{\omega^2-{\omega'}^2}\\\notag
&+4J^m(\omega,\ell)\int_0^\infty  d\omega'J^k(\omega',\ell)
\frac{\omega\omega'}{\omega^2-{\omega'}^2}
\end{align}
We defined $\bar k=(z,y)$ for $k=(y,z)$ and $J^m(\omega,\ell)\equiv\sum\al\lambda\al^y\lambda\al^z\delta(\omega-\omega\al)=(J^y(\omega,\ell)J^z(\omega,\ell))^{1/2}$.

The truncated flow of the observable shall be given by $\sz(\ell)=h(\ell)\sz+\sx\sum\al\chi\al(\ell)(b\al+b\al\dag)$. The flow equations $\partial_\ell\sz=[\eta,\sz]$ are subjected to the same decoupling procedure as above. 
Defining the spectral functions $S^k(\omega,\ell)\equiv\sum\al\lambda\al^k(\ell)\chi\al(\ell)\delta(\omega-\omega\al)$, we obtain the following flow equations:
\begin{align}
\label{FlowEquations_Extended_SpinBoson_Observable}
\dl h(\ell) &=-\int_0^\infty  d\omega \omega S^y(\omega,\ell)\quad,
\end{align}
\begin{widetext}
\begin{align}
\begin{split}
\dl S^y(\omega,\ell)&=h(\ell)\omega J^y(\omega,\ell)-\omega^2S^y(\omega,\ell)+\De(\ell)\omega S^z(\omega,\ell)\\
&+J^m(\omega,\ell)\int_0^\infty  d\omega'S^y(\omega',\ell)
\frac{\omega^2+{\omega'}^2}{\omega^2-{\omega'}^2}
+2J^y(\omega,\ell)\int_0^\infty  d\omega'S^z(\omega',\ell)
\frac{\omega\omega'}{\omega^2-{\omega'}^2}\\
&+S^z(\omega,\ell)\int_0^\infty  d\omega'J^y(\omega',\ell)
\frac{\omega^2+{\omega'}^2}{\omega^2-{\omega'}^2}
+2S^y(\omega,\ell)\int_0^\infty  d\omega'J^m(\omega',\ell)
\frac{\omega\omega'}{\omega^2-{\omega'}^2}
\end{split}
\end{align}
\end{widetext}
Since $S^z(\omega,\ell)=S^y(\omega,\ell)J^z(\omega,\ell)/J^m(\omega,\ell)$, it suffices to define the integro-differential equation for $S^y(\omega,\ell)$.

One criterion for the assessment of the quality of the flow equations is again given by $\langle \sz^2(\ell)\rangle$ which should be approximately equal to one for all $\ell$. It is governed by the differential equation 
\begin{align}
\partial_\ell\langle \sz^2(\ell)\rangle=-\int_0^\infty d\omega  d\omega'\frac{\omega-{\omega'}}{\omega+{\omega'}}S^y(\omega,\ell)S^z(\omega',\ell).
\end{align}
In the asymptotic regime $\ell\to\infty$, the flow equations simplify considerably. i.e. $S^y(\omega,\ell)=S^z(\omega,\ell)$ which will be shown below. This property guarantees that the spectral weight of correlation functions is asymptotically conserved - in contrary to the form-invariant flow. Numerical calculations show that - throughout the flow - the spectral sum rule is fulfilled within less than $0.1\%$ relative error. This is a significant improvement w.r.t. the form-invariant flow.

\subsubsection{Asymptotic behavior} 
We now discuss the asymptotic behavior.
As a first approximation, we take $\lambda\al^y=\lambda\al^z$ and thus drop the index $k$ of the spectral function. This assumption is consistent with the flow equations and justified later based on the numerical result.
The asymptotic flow equations are then solved by the ansatz 
\begin{align}
\notag
\De(\ell)\to a\ell^{-1/2}\quad,\quad
J(\omega,\ell)\to\ell^{-1/2}J(\omega\sqrt{\ell})
\end{align}
which leads to the following non-local differential equation for $J(y)$ with the scale-invariant variable $y\equiv\omega\sqrt{\ell}$:
\begin{align}
\partial_y J(y)&=-4J(y)\Big((y-a)-2\int_0^\infty dy'\frac{J(y')}{y-y'}-\frac{t}{4y}\Big),\label{AsympJEx}
\end{align}
where $t\equiv1-4\int_0^\infty dyJ(y)$. The spectral function depends on the parameter $a$  and the boundary condition is given by $a=4\int_0^\infty dyyJ(y)$.

With the similar ansatz for the observable flow, i.e. 
\begin{align}
\notag
h(\ell)=b\ell^{-1/4-s'/4}\;,\; S(\omega,\ell)\to\ell^{-1/4-s'/4}S(\omega\sqrt{\ell})\;,
\end{align}
we obtain the following non-local differential equation for $S(y)$ with $y=\omega\sqrt{\ell}$:
\begin{align}
\notag
\partial_y S(y)&=-2S(y)\Big((y-a)-2\int_0^\infty dy'\frac{J(y')}{y-y'}-\frac{t'}{2y}\Big)\notag\\\label{AsympSEx}
&+2J(y)\Big(b+2\int_0^\infty dy'\frac{S(y')}{y-y'}+\frac{t''}{y}\Big)\quad,
\end{align}
where $t'\equiv(s'+t)/2$ and $t''\equiv\int_0^\infty dyS(y)$. The parameter $b$ is given by $b=-4\int_0^\infty dyyS(y)/(s'+1)$. 

\begin{figure}[t]
  \begin{center}
    \epsfig{file=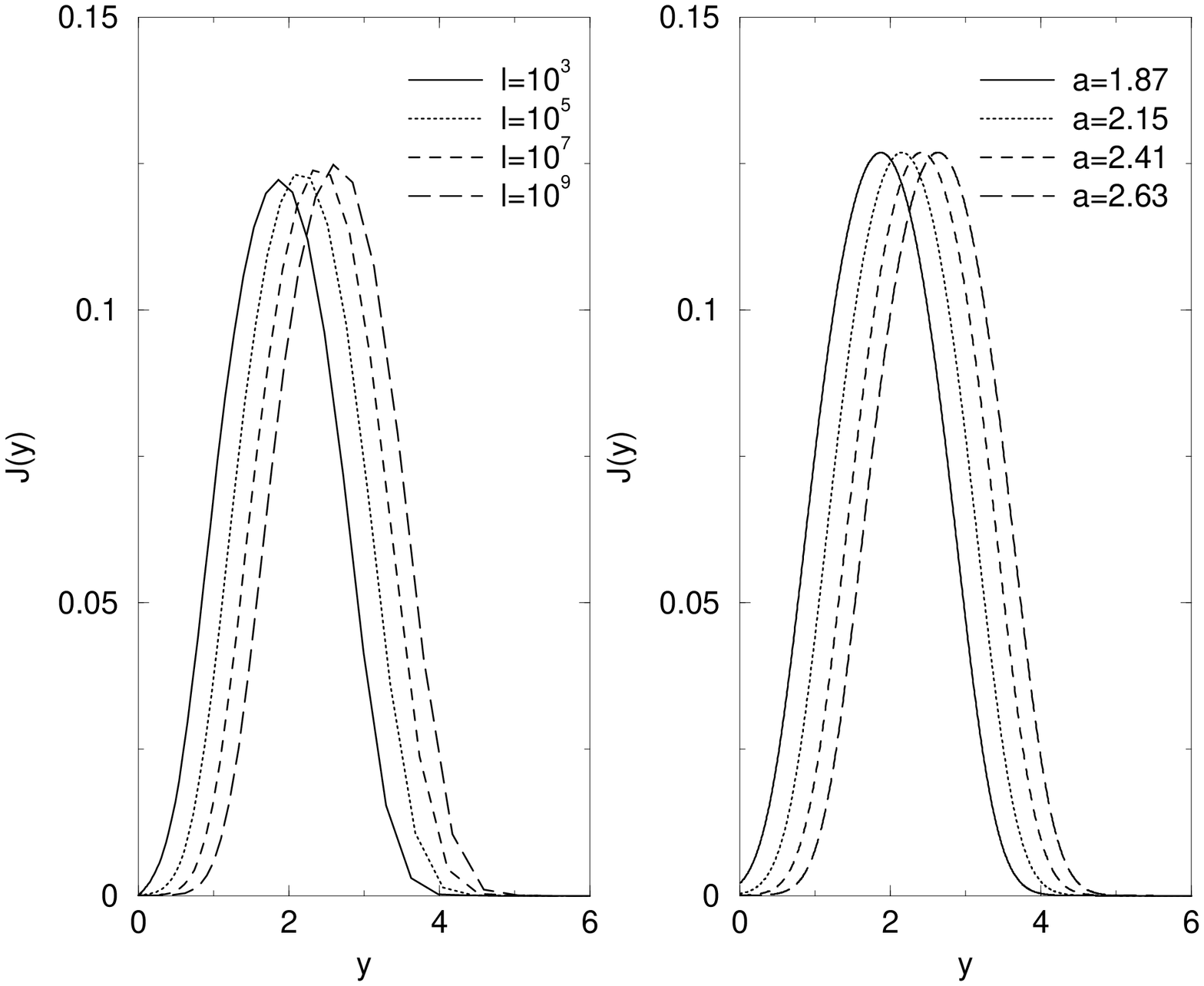,height=7.5cm,angle=-90}
    \caption{Left hand side: The asymptotic spectral functions $J(y)$ obtained from the extended flow for Ohmic coupling with $\alpha=0.1$ and $\omega_c/\De_0=10$ at various values of $\ell$. Right hand side: The ``analytic'' solution following from Eq. (\ref{nonlocalJSB}) for the corresponding values of $a(\ell)=\De(\ell)\ell^{1/2}$ as they follow from the flow equations. A numerical fit suggests $a\to4.5$ for $\ell\to\infty$.}
    \label{AsympJExOhmic}
  \end{center}
\end{figure}
\begin{figure}[t]
  \begin{center}
    \epsfig{file=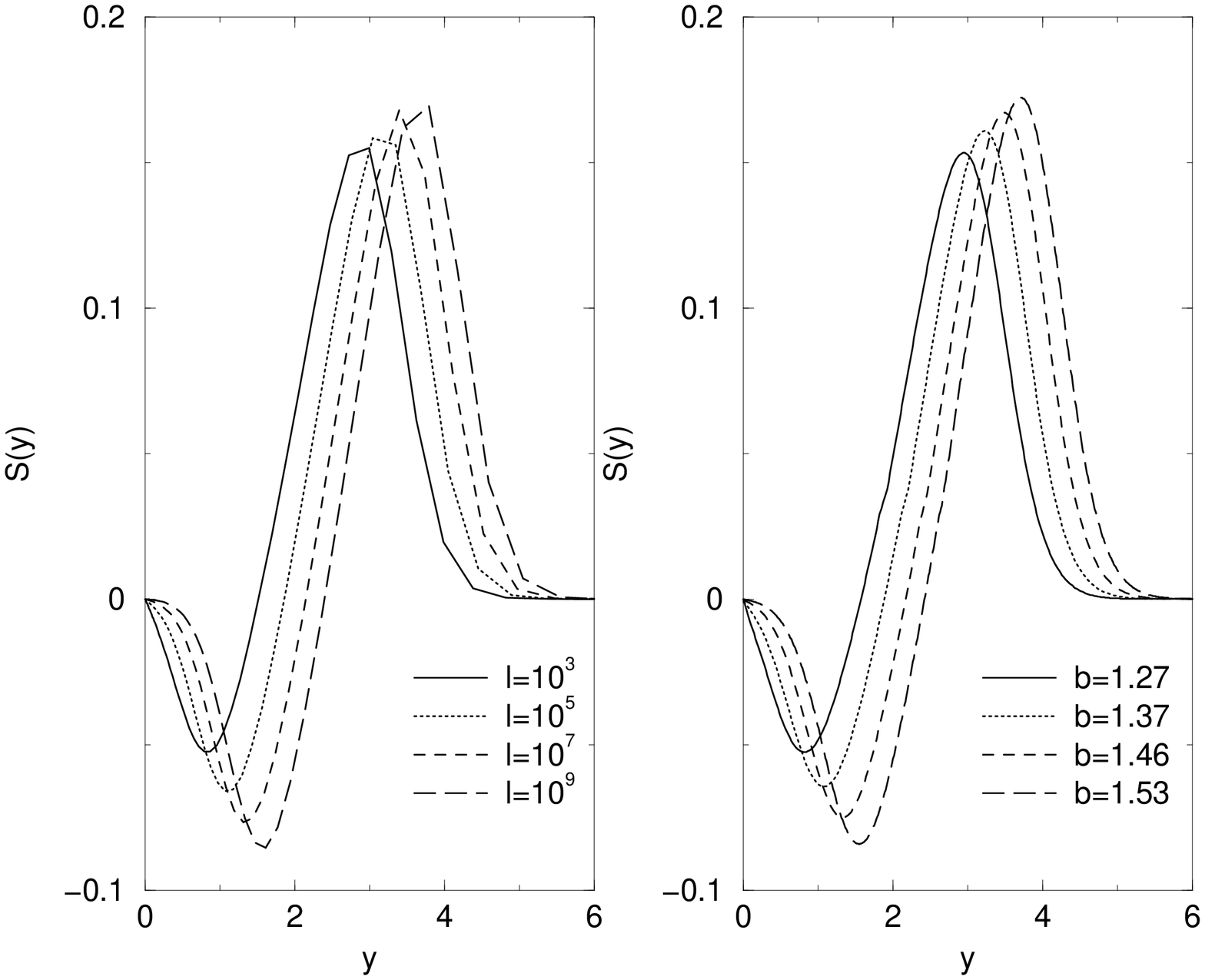,height=7.5cm,angle=-90}
    \caption{Left hand side: The asymptotic spectral functions $S(y)$ obtained from the extended flow for Ohmic coupling with $\alpha=0.1$ and $\omega_c/\De_0=10$ at various values of $\ell$. Right hand side: The ``analytic'' solution following from Eq. (\ref{Asymp_wwSEx}) for the corresponding values of $b(\ell)=h(\ell)\ell^{1/2}$ as they follow from the flow equations. A numerical fit suggests $b\to2.0$ for $\ell\to\infty$.}
    \label{AsympSExOhmic}
  \end{center}
\end{figure}

The above ansatz assumes that $a=\De(\ell)\ell^{1/2}$ is constant. This is not the case, though - for the entire regime that is accessible by our numerical integration. We thus extend the asymptotic behavior of the tunnel-matrix element $\De(\ell)$ by setting $\De(\ell)=a(\ell)\ell^{-1/2}$ with $a(\ell)\to a$. The generalization induces a shift in the spectral function, i.e. $J(\omega,\ell)\to \ell^{-1/2}J(\omega\sqrt{\ell}+\hat{a}(\ell))$ with $[a-a(\ell)]/\hat a(\ell)=$ const. 

The same extensions are also necessary in the case of the asymptotic observable flow, i.e. $h(\ell)\to b(\ell)\ell^{-1/4-s'/4}$ with $b(\ell)\to b$ and $S(\omega,\ell)\to\ell^{-1/4-s'/4}S(\omega\sqrt\ell+\hat a(\ell))$ with $[b-b(\ell)]/\hat a(\ell)=$ const. All limits hold for $\ell\to\infty$. A numerical fit for Ohmic dissipation with $\alpha=0.1$ and $\omega_c/\De_0=10$ - based on the resulting differential equation for $a(\ell)$ and $b(\ell)$ and on the fact that $[a-a(\ell)]/[b-b(\ell)]=$ const - suggests that $a=4.5$ and $b=2.0$. 

But in the following, we want to demonstrate that the $\ell$-dependent shift in the spectral function does not essentially alter the functional dependence of $J(y)$ and $S(y)$. For this, we define $\widetilde J(y)\equiv J(y+a)$. Substituting $y\to y+a$ in Eq. (\ref{AsympJEx}) and setting $a\to\infty$ on the lower bound of the integral, the boundary condition yields $t=0$ and we obtain
\begin{align}
\label{nonlocalJSB}
\partial_y \widetilde J(y)&=-4y\widetilde J(y)\Big(1-4\int_0^{\infty}dy'\frac{\widetilde J(y')}{y^2-{y'}^2}\Big).
\end{align}  
This is the same non-local differential equation as in the case of the dissipative harmonic oscillator and the spin-boson model with form-invariant flow. But this time the parameter $a$ of Eq. (\ref{nonlocalJHarm}) is given by $a=2$. The ``analytic'' result of the above equation is in good agreement with the flow equation result after a shift by the corresponding $a(\ell)$. This is demonstrated in Fig. \ref{AsympJExOhmic}, where on the lhs the flow equations were integrated numerically and on the rhs Eq. (\ref{nonlocalJSB}) was used. The initial spectral function was taken to be $J(\omega)=2\alpha\omega\Theta(\omega_c-\omega)$ with $\alpha=0.1$ and $\omega_c/\De_0=10$ and the various values of $a(\ell)$ were obtained from the numerical integration of the flow equations.

The same transformation can be done for the asymptotic function of the observable. With $\widetilde S(y)\equiv S(y+a)$ this yields the following equation:
\begin{align}
\notag
\partial_y \widetilde S(y)&=-2\widetilde S(y)\Big(y-2\int_{-\infty}^\infty dy'\frac{\widetilde J(y')}{y-y'}+\frac{s'}{4(y+a)}\Big)\\
&+2\widetilde J(y)\Big(b+2\int_{-\infty}^\infty dy'\frac{\widetilde S(y')}{y-y'}+\frac{t''}{y+a}\Big)\quad\label{Asymp_wwSEx}
\end{align}
With $4t''=-s'b/a$ which follows from the boundary condition, the ``analytic'' solutions are again in good agreement with the flow equation solutions after a shift by the corresponding $a(\ell)$. This is demonstrated in Fig. \ref{AsympSExOhmic}, where on the lhs the flow equations were integrated numerically and on the rhs Eq. (\ref{Asymp_wwSEx}) was used. The various values of $b(\ell)$ were obtained from the numerical integration of the flow equations.

Keeping in mind that $J(y)$ is exponentially small for $y\to0$, we are allowed to neglect the inhomogeneous contribution in Eq. (\ref{AsympSEx}) or (\ref{Asymp_wwSEx}) in this regime. The asymptotic behavior of $S(y)$ is then given by $S^2(y)\propto J(y)y^{s'}$ for $y\to0$. With $J(y)\to J_0$ and $S(y)\to S_0$ for $y\to0$, we thus obtain the $\ell$-independent correlation function for $\omega\to0$
\begin{align}
C(\omega\to0)=\omega^{s'}\frac{S_0^2}{J_0}\quad.
\end{align}
For an initial spectral function $J(\omega)\propto \omega^s$, $s=s'$ is verified by the numerical data and we thus obtain the correct low-energy behavior $C(\omega)\propto\omega^s$.

We want to note that the equations for the asymptotic behavior hold independent of the initial coupling strength and of the bath type. The initial conditions only enter via the parameters $a$ and $b$ for which the corresponding asymptotic functions are obtained. But we also want to mention that for higher coupling strength $\alpha$ and for lower values of $s$ the asymptotic regime $\lambda\al^y=\lambda\al^z$ is only reached for rather large $\ell$. For an initial spectral function with $\alpha=0.1$, $\omega_c/\De_0=10$, $K/\De_0=1$ and $s=0.3$ we have e.g. $\De_0^2\ell^*\gapp10^{20}$, $\ell^*$ denoting the asymptotic regime were $\lambda\al^y=\lambda\al^z$ holds on all energy scales. The corresponding parameters at $\ell^*$ are given by $a=2.23$ and $b=1.43$. 

Nevertheless, one can extend the range of the asymptotic regime if one drops the assumption $\lambda\al^y=\lambda\al^z$, i.e. one makes the ansatz $J^k(\omega,\ell)\to\ell^{-1/2}J^k(\omega\sqrt{\ell})$ and $S^k(\omega,\ell)\to\ell^{-1/4-s'/4}S^k(\omega\sqrt{\ell})$ for $k=y,z$ . The resulting set of non-local differential equations for the functions $J^k(y)$ recovers the flow equation results also for $\tilde\ell^*\lapp\ell^*$, if the second boundary condition is suitably chosen. In fact, $\tilde\ell^*$ can be chosen such that $\tilde\ell^*\ll\ell^*$. 

To conclude the discussion, we now derive the asymptotic conservation law, in analogy to Eq. (\ref{K_a_Asmptotic}). For this, we define the $\ell$-dependent correlation function $C(\omega,\ell)\equiv\sum_\alpha\chi\al^2(\ell)\delta(\omega-\omega\al)$ and the asymptotic spectral function $C_a(\omega,\ell)\equiv C(\omega)-C(\omega,\ell)$. For $\tilde\ell^*$ being in the quasi-asymptotic regime, we obtain from the above flow equations $C_a(\omega,\tilde\ell^*)=\omega^{s'}I( \tilde y^*)$ where $I( \tilde y^*)$ is given by 
\begin{widetext}
\begin{align}
\label{FinalAsymptotics}
I( \tilde y^*)\equiv4\int_{ \tilde y^*}^\infty dy y^{-s'}\Big(bS^y(y)+S^z(y)\int_0^\infty dy'S^y(y')G^+(y,y')+S^y(y)\int_0^\infty dy'S^z(y')G^-(y,y')+t^{''}\frac{S^z(y)}{y}\Big).
\end{align}
\end{widetext} 
We defined $G^\pm(y,y')\equiv(y-y')^{-1}\pm(y+y')^{-1}$ and $ \tilde y^*=\omega\sqrt{\tilde\ell^*}$. The above expression simplifies considerably for $\ell^*$ being in the asymptotic regime.\cite{Sta02} 
 
\subsubsection{Numerical results}

We will now present the numerical results for the equilibrium correlation function $C(\omega)$.  

\begin{figure}[t]
  \begin{center}
    \epsfig{file=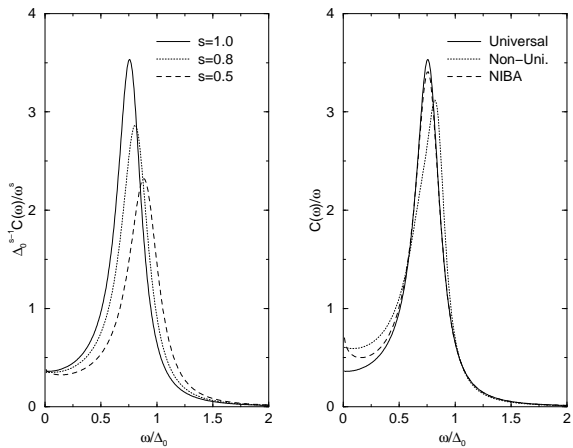,height=7.5cm,angle=-90}
    \caption{Left hand side: The spectral function $C(\omega)/\omega^s$ for $J(\omega)=2\alpha K^{1-s}\omega^s\Theta(\omega_c-\omega)$ with $\alpha=0.1$, $K/\De_0=1$, $\omega_c/\De_0=10$ for various bath types $s$. Right hand side: The spectral function $C(\omega)/\omega$ for $J(\omega)=2\alpha\omega\Theta(\omega_c-\omega)$ with $\alpha=0.1$ and $\omega_c/\De_0=10$ as they follow from universal (solid line) and non-universal (dotted line) asymptotic behavior (for details, see text). The NIBA result is shown as dashed line.}
    \label{FinalFinalFig}
  \end{center}
\end{figure}

On the left hand side of Fig. \ref{FinalFinalFig}, the spectral function $C(\omega)/\omega^s$ is shown for various bath types $s$. The initial coupling function was chosen to be $J(\omega)=2\alpha K^{1-s}\omega^s\Theta(\omega_c-\omega)$ with $\alpha=0.1$, $K/\De_0=1$, and $\omega_c/\De_0=10$. As can be seen, $C(\omega)/\omega^s$ is constant for low energies. But notice that $\partial_\omega C(\omega)/\omega^s\neq0$ for $\omega\to0$. This point was already mentioned in the discussion of the dissipative harmonic oscillator (see Fig. \ref{Abschluss_s}).  

Apart from the sum rule $\langle \sigma_z(\ell)\rangle\approx1$ for all $\ell$, another criterion for the assessment of the quality of the flow equations is given by the generalized Shiba relation:\cite{Sas90} 
\begin{align}
\label{Shiba}
\lim_{\omega\to0}\frac{C(\omega)}{\omega^s}=2\alpha K^{1-s}\Big(\int_0^\infty d\omega\frac{C(\omega)}{\omega}\Big)^2
\end{align} 
The results are given in Table \ref{TableShiba}.
So far, to our knowledge, the generalized Shiba relation for sub-Ohmic dissipation can only be checked by the above procedure. 

For Ohmic dissipation, the outlined approach can be contrasted with previous results. As initial spectral function we use  $J(\omega)=2\alpha\omega\Theta(\omega_c-\omega)$ with $\alpha=0.1$ and $\omega_c/\De_0=10$. On the right hand side of Fig. \ref{FinalFinalFig}, the spectral function $C(\omega)/\omega$ is shown in comparison to the result obtained by integrating the flow equations of Sec. IIIA with $f(\omega,\ell)=-(\omega-\De(\ell))/(\omega+\De(\ell))$. It is obtained by integrating the flow equations up to $\De_0^2\ell^*=100$ and then using the conservation law of Eq. (\ref{ConservationHarm}). For higher values of $\De_0^2\ell^*\gapp100$, the spectral function shows instabilities for energies close to the resonance $\omega\approx\De^*$. The situation becomes worse for higher values of $\alpha$ and generally for sub-Ohmic baths. For a detailed discussion, see Ref. \onlinecite{Keh97}. 

In Table \ref{ContrastShiba}, the two flow equation approaches are contrasted with the help of the generalized Shiba relation for various weak Ohmic couplings. Throughout the parameter regime, flow equations with universal asymptotic behavior yield better results. For $\alpha=0.2$, the non-universal approach yields no stable result in order to check the generalized Shiba relation.

On the right hand side of Fig. \ref{FinalFinalFig}, $C(\omega)/\omega$ is also shown as obtained from the non-interacting blib approximation (NIBA) for suitable effective tunneling $\Delta_{\text{eff}}$.\cite{Leg87} The curve coincides with the one obtained from the universal approach for short and intermediate time scales. For long time scales, the graph diverges as $C(\omega)/\omega\propto\omega^{-2\alpha}$, which is incorrect.    

The universal flow equation approach is applicable for not too small bath types $s\geq s^*$ and not too large coupling strengths $\alpha\leq\alpha^*$. As criterion we again use the generalized Shiba relation and define the ratio $R\equiv[\lim_{\omega\to0}C(\omega)/\omega^s]/[2\alpha K^{1-s}(\int_0^\infty d\omega C(\omega)/\omega)^2]$. We demand that $|1-R|\lapp0.3$.
For Ohmic coupling and $\omega_c/\De_0=10$, e.g., we have $\alpha^*\lapp0.4$ and for a fixed coupling strength, $\alpha=0.1$ say, we have $s^*\gapp0.3$ ($K/\De_0=1$). These parameter regimes are further discussed in Ref. \onlinecite{Sta02} where also the case of finite bias is treated. 

If one is only interested in the effective tunnel-matrix element or the half-width of the resonance, the flow equations with universal asymptotic behavior can yield the correct result even if the Shiba relation is not fulfilled satisfactorily.
\begin{table}
\begin{tabular}{c||c|c|c}
&$\lim_{\omega\to0}C(\omega)/\omega^s$&$2\alpha K^{1-s}(\int_0^\infty d\omega C(\omega)/\omega)^2$&Ratio\\
\hline
s=1.0&0.364&0.391&0.93\\
s=0.8&0.356&0.390&0.91\\
s=0.5&0.384&0.432&0.89
\end{tabular}
\caption{The generalized Shiba relation for various bath types $s$ with $\alpha=0.1$, $K/\De_0=1$, and $\omega_c/\De_0=10$.}
\label{TableShiba}
\end{table}
\begin{table}
\begin{tabular}{c||c|c|c|c}
R&$\alpha=0.01$&$\alpha=0.05$&$\alpha=0.1$&$\alpha=0.2$\\
\hline   
Universal&1.00&0.97&0.93&0.86\\
Non-Uni.&1.03&1.19&1.38&-
\end{tabular}
\caption{Contrasting the universal and non-universal flow equations with the help of the generalized Shiba relation, $R\equiv[\lim_{\omega\to0}C(\omega)/\omega]/[2\alpha (\int_0^\infty d\omega C(\omega)/\omega)^2]$, for a Ohmic bath with $\omega_c/\De_0=10$ and various coupling strengths $\alpha$.}
\label{ContrastShiba}
\end{table}
\section{Conclusions}
In this work, we applied the flow equation method to two dissipative systems, the exactly solvable dissipative harmonic oscillator and the spin-boson model. 
Investigating the former model, we have pointed that there are two basic universality classes for flow equations, i.e. either $\omega_{\infty}>0$ and Eq. (\ref{FreqSelfCon}) holds or $\omega_{\infty}=0$. The former class was labeled non-universal since the asymptotic behavior depended on the initial conditions of the system. The latter class was labeled universal since the fixed point Hamiltonian was given by the non-interacting bath, only. The universal asymptotic behavior of the spectral coupling function as well as of the observable are characterized by scale-invariant, non-local differential equations from which the low-energy behavior of correlation functions can be deduced analytically.

These ideas were also applied to the non-trivial spin-boson model. It was mentioned that the form-invariant flow of the Hamiltonian necessarily displays universal asymptotic behavior when the reflection symmetry is broken. The resulting flow equations, though, did not yield the correct normalization condition nor the correct low-energy behavior of correlation functions. Nevertheless, an effective energy scale was recovered which was in good agreement with the predictions of the NIBA. 
 
Due to the obvious shortcomings of the form-invariant flow, we set up extended flow equations which also displayed universal asymptotic behavior. The resulting flow equations yielded very good results for a wide range of the parameter regime where also sub-Ohmic baths could be included in the treatment. We further deduced the correct low-energy behavior of correlation functions from the universal asymptotic behavior and showed that the normalization condition and the Shiba relations were satisfied within numerical errors for a certain parameter regime.

Universal asymptotic behavior is crucial in order to determine correlation functions within the flow equation approach since only then one can assure that the dominant weight of the correlation functions is accounted for by the stable intermediate flow. Apart from this technical preference, the trivial fixed point Hamiltonian also accounts for the correct spectrum which is not superposed by the spectrum of the isolated system. We also want to mention that the asymptotic flow of the spectral function is described by the same non-local differential equation - independent of the model or the truncation scheme, but only differing by one parameter. 

The presented procedure can easily be extended to other dissipative models by starting from the canonical generator $\eta_c=[H_B,H]$ where $H_B$ denotes the non-interacting bath.\\ 

It is a pleasure to thank A. Mielke for innumerable discussions on the flow equation method. This work was funded by the Deutsche Forschungsgesellschaft.


\begin{thebibliography}{99}
\bibitem{FootEne}
The flow parameter $\ell$ has dimensions of inverse energy squared if one chooses the generator as $\eta=[H_0,H]$, with $H$ denoting the full and $H_0$ the diagonal Hamiltonian.
\bibitem{FootKM}
The mapping is only successful for super-Ohmic baths and for Ohmic baths at low coupling.
\bibitem{FootBS} 
We explicitly exclude the possibility of a quantum phase transition. Our approach will therefore only be valid for weak coupling.
\bibitem{FootScales}
We have chosen this parametrization to emphasize that the approach can easily be extended to more general, e.g. periodic, potentials, see Ref. \onlinecite{StaD}.
\bibitem{FootZT}
Throughout this work, we consider the zero temperature limit, $T=0$.
\bibitem{FootDrop}
Notice that we dropped the index $s'$ on the functions $J(\omega,\ell)$, $\hat{J}(\ell)$, and $\bar{J}(y)$.
\bibitem{FootFS}
Again, we neglect contributions of order $1/N$, $N$ being the number of modes.
\end{thebibliography}
\end{document}